\documentclass[conference]{IEEEtran}
\IEEEoverridecommandlockouts
\usepackage[utf8]{inputenc} 
\usepackage[T1]{fontenc}
\usepackage{url}
\usepackage{ifthen}
\usepackage{cite}
\usepackage[cmex10]{amsmath} % Use the [cmex10] option to ensure complicance
\usepackage{xparse,amssymb,mathtools,color,graphicx,subcaption,algpseudocode,algorithm,comment,dsfont,amsthm,color}

\newtheorem{theorem}{Theorem}
\newtheorem{lemma}{Lemma}
\newtheorem{definition}{Definition}

\newcommand{\Z}{\mathbb{Z}}
\newcommand{\R}{\mathbb{R}}

\newcommand{\cR}{{\cal R}}

\NewDocumentCommand\sqn{mg}{%
    \|\mathbf{#1}_{\IfNoValueTF{#2}{}{#2}}\|^2%
}

\def\b{\mathbf}

\graphicspath{{../main_document/Figures/}} % Specifies the directory where pictures are stored
\DeclareGraphicsExtensions{.pdf,.jpeg,.png,.jpg}
\DeclareMathOperator*{\argmin}{arg\,min}
\DeclareMathOperator*{\argmax}{arg\,max}
\newcommand\Algphase[1]{%
\vspace*{-.7\baselineskip}\Statex\hspace*{\dimexpr-\algorithmicindent-2pt\relax}\rule{0.45\textwidth}{0.4pt}%
\Statex\hspace*{-\algorithmicindent}\textbf{#1}%
\vspace*{-.7\baselineskip}\Statex\hspace*{\dimexpr-\algorithmicindent-2pt\relax}\rule{0.45\textwidth}{0.4pt}%
}
\DeclareMathOperator{\EX}{\mathbb{E}}% expected value

\begin{document}
\title{Asymptotically Optimal Scheduling for Compute-and-Forward 
\thanks{This research was partially supported by European Unions Horizon 2020
Research and Innovation Program SUPERFLUIDITY, Grant Agreement
671566.}
} 

% %%% Single author, or several authors with same affiliation:
% \author{%
%   \IEEEauthorblockN{Stefan M.~Moser}
%   \IEEEauthorblockA{ETH Zürich\\
%                     ISI (D-ITET)\\
%                     CH-8092 Zürich, Switzerland\\
%                     Email: moser@isi.ee.ethz.ch}
% }
\author{\IEEEauthorblockN{Ori Shmuel, Asaf Cohen, Omer Gurewitz}
\IEEEauthorblockA{Ben-Gurion University of the Negev,\\ \{shmuelor, coasaf, gurewitz\}@bgu.ac.il}
}

\maketitle

%%%%%%
%% Abstract: 
%% If your paper is eligible for the student paper award, please add
%% the comment "THIS PAPER IS ELIGIBLE FOR THE STUDENT PAPER
%% AWARD." as a first line in the abstract. 
%% For the final version of the accepted paper, please do not forget
%% to remove this comment!
%%
\begin{abstract}
Consider a Compute and Forward (CF) relay network with $L$ users and a single relay. The relay tries to decode a linear function of the transmitted signals. For such a network, letting all $L$ users transmit simultaneously, especially when $L$ is large, causes a significant degradation in the rate in which the relay is able to decode. In fact, the rate goes to zero very fast with $L$. Therefore, in each transmission phase only a fixed number of users should transmit, i.e., users should be scheduled. 
  
  In this work, we examine the problem of scheduling for CF and lay the foundations for identifying the optimal schedule which, to date, lacks a clear understanding. Specifically, we start with insights why when the number of users is large, good scheduling opportunities can be found.  Then, we provide an asymptotically optimal, polynomial time scheduling algorithm and analyze it's performance. We conclude that scheduling under CF provides a gain in the system sum-rate, up to the optimal scaling law of $O(\log{\log{L}})$.

  \end{abstract}

%To cope with the computational complexity of choosing the optimal subset of transmitters, and the optimal $\b{a}$ for that subset, we first devise a sub-optimal algorithm, which finds the best subset for a given, fixed vector $\b{a}$ in polynomial time. We then prove that, in fact, this choice of $\b{a}$ is asymptotically optimal, in the sense that it achieves the optimal scaling law.

%%%%%%%%%%%%%%%%%%%%%%%%%%%%%%%%%%%%%%%%%%%%%%%%%%%%%%%%%%%%%%%%%%%%%%%%%%%%%%%%%%%%%%

%% The paper must be self-contained. However, if you are referring to
%% a full version for checking certain proofs, please provide the
%% publically accessible location below.  If the paper is completely
%% self-contained, you can remove the following line from your
%% submission.
%\textit{A full version of this paper is accessible at: \url{http://www.isit2018.org/}}

\section{Introduction}
Compute and Forward (CF) is a coding strategy \cite{nazer2011compute}, introduced for relay systems consisting of multiple transmitters and multiple relays. In this scheme, the relays (receivers) decode a linear function of the received messages instead of decoding them individually. Therefore, it is a powerful technique for mitigating users' interferences, which is a prominent problem in today's wireless communication systems. The ability to exploit simultaneous transmissions is possible due to the usage of lattice codes, which enable the function of the transmitted messages to be decoded as a legitimate message. Accordingly, in recent years, the main ideas of CF were used to attain new results for linear receivers \cite{zhan2014integer}, the Multiple Access Channel sum capacity \cite{ordentlich2014approximate} and more \cite{wei2012compute,hong2013compute}. 

The performance of CF in the regime of large number of users and a fixed number of relays was investigated in \cite{shmuel2017necessity}. It was shown that the CF scheme degenerates fast when the number of transmitters grows, in the sense that the relays prefer to decode a single message instead of any other linear combination of the messages. This is due to the "self" noise added to the decoding process, which tries to approximate the real channel vector with an integer coefficients vector. Specifically, as the number of simultaneously transmitting users grows, a receiver essentially prefers to treat all messages as noise apart from the message it tries to decode. As a consequence, the system's sum-rate goes to zero (even for a moderate number of users) and in order to have any guarantee for non-zero rate, user scheduling must be applied. Namely, applying CF on large scale relaying systems, where there is a fixed number of relays, without a restriction on the number of transmitting users, would be futile.

On the other hand, restricting the number of transmitting users can provide coding opportunities for the CF scheduler, by scheduling users with channel conditions which are more favorable for CF while grouped together, providing gain to the overall system's sum-rate. A related improvement, while scheduling in CF networks, was presented also in \cite{ramirez2015scheduling}. Therein, the authors showed by simulation that even a simple scheduling scheme can be useful. However, no performance guarantees or analysis for the optimal schedule were carried out. In this work, we examine user scheduling in the context of CF schemes and explore the scheduling considerations a CF scheduler should take. Specifically, as CF relies on the appropriate match between the fading coefficients of the transmitted signals and a certain linear function with integer coefficients, with scheduling, one can influence not only the signals which participate, but also the proper choice of the function. 

\subsubsection*{Main contributions}
We consider a simple relay network with a single relay and $L$ transmitters where, due to the necessity for restricting the number of simultaneously transmitting users, we schedule $k$ users for transmission. This setting is sufficient to attain important results and insights for scheduling in CF, and build the first steps for comprehending what is the optimal schedule for CF networks. 

We begin with the analysis of the optimal schedule and present an asymptotically optimal, polynomial time scheduling algorithm for CF, which maximizes the system sum-rate. The algorithm is analyzed, and its performance is lower and upper bounded. The lower bound is derived using probabilistic arguments on the properties of the optimal schedule and the upper bound is derived using a universal upper bound on the performances of CF. We show that both the lower bound and the upper bound scale as $O(\log{\log{L}})$, which essentially proves the optimality of the suggested algorithm and the specific schedule it provides. Consequently, we are able to provide an important property of the optimal schedule; the scheduler will seek groups of users which best match a fixed, non-trivial yet deterministic coefficients vector. Therefore, we show that the gain arises solely from the proper choice of users for that vector, and not from actually optimizing on the coefficients vector, like CF suggests for finite systems.

%%%%%%%%%%%%%%%%%%%%%%%%%%%%%%%%%%%%%%%%%%%%%%%%%%%%%%%%%%%%%%%%%%%%%%%%%%%%%%%%%%%%%%

\section{System Model and Known Results} \label{sec-System model}
Consider a multi-user, single-relay network, where there are $L$ users and a single relay. Each transmitter sends a real-valued codeword, $\mathbf{x}_l\in \R^n$ with rate $R$, which is subject to a power constraint $P$. 
The relay observes a noisy linear combination of the transmitted signals through the channel,
\begin{equation}
\b{y}=\sum_{l=1}^{L}h_{l}\mathbf{x}_l+\b{z}, 
\end{equation}
where $h_{l} \sim \mathcal{N}(0,1), \ 1\leq l \leq L$, are the real channel coefficients and $\b{z}$ is an i.i.d., Gaussian noise, $\b{z} \sim \mathcal{N}(0,\b{I})$. Let $\b{h}_L= [h_{1},h_{2},...,h_{L}]^T$ denote the vector of channel coefficients of all transmitting users. We assume that the relay knows the channel vector $\b{h}_L$. In CF, after receiving the noisy linear combination, the relay selects an integer coefficients vector $\b{a}=(a_{1},a_{2},...,a_{L})^T \in \mathbb{Z}^L$, and attempts to decode the lattice point $\sum_{l=1}^L a_l\b{x}_l$ from $\b{y}$. It then forward the decoded code-word towards the destination via a dedicated channel or another relay. The decoder, upon receiving enough such linear combinations of messages, decodes the original messages by solving the system of linear equations obtained from the coefficients vectors and the received code-words in each transmission phase. That is, successful decoding at the decoder is conditioned on the ability of the relay to decode the correct linear combination of messages and the full rank (i.e., rank $L$) of the matrix $\b{A}$ which its raws are the coefficients vectors of each phase. Accordingly, we note that the number of transmission phases may be more than $L$. 

The computation rate of the linear combination, with respect to the coefficients vector $\b{a}$, is \cite{nazer2011compute}:
\begin{equation}\label{equ-Computation rate with MMSE}
\cR(\b{h}_L,\b{a})= \frac{1}{2} \log^+ \left(\left( \|\b{a}\|^2- \frac{P(\b{h}_L^T\b{a})^2}{1+P\|\b{h}_L\|^2} \right)^{-1}\right),
\end{equation}
where $\log^+(x)\triangleq \max \{\log(x),0\}$.

In order for the relay to decode a linear combination with a coefficients vector $\b{a}$, all messages rates, for messages which have a non-zero value in their corresponding entry in $\b{a}$, must comply with the rate in \eqref{equ-Computation rate with MMSE}. That is,  $R< \cR(\b{h}_L,\b{a})$. Note that only messages with a non-zero entry in $\b{a}$ are considered in the linear combination. We thus define as our performance measure the system's sum of computation rates to be the number of non-zero entries in $\b{a}$ times $\cR(\b{h}_L,\b{a})$. This metric captures the computation rate of the relay along with the number of messages which are not considered as noise and take an active part in the decoding of the linear combination.

Since the relay can decide which linear combination to decode (i.e., to choose the coefficients vector $\b{a}$), the relay can choose $\b{a}$ which maximizes $\cR(\b{h}_L,\b{a})$ for a given $\b{h}_L$. Note that according to \cite[Lemma 1]{nazer2011compute}, the search domain for this maximizing $\b{a}$ is restricted to all vectors $\b{a}$ for which $\sqn{a} \leq 1+P\sqn{h}{L}$. A polynomial time algorithm with complexity $O(L^2\sqrt{1+P\sqn{h}{L}})$, which finds this maximizing $\b{a}$, was introduced in \cite{sahraei2014compute}.

When the number of users $L$ is large, in \cite{shmuel2017necessity} we showed that with probability that goes to 1 with $L$, the coefficients vector which will maximize $\cR(\b{h}_L,\b{a})$ is actually a unit vector. Specifically, \cite{shmuel2017necessity} provides the following result.

\begin{theorem}[\cite{shmuel2017necessity}]\label{the-Probability for having a unit vector as the maximaizer}
\textit{Under the CF scheme, the probability that a non-trivial vector $\b{a}$ will be the coefficients vector which maximize the achievable rate $\mathcal{R}(\b{h}_L,\b{a})$, compared to a unit vector $\b{e}_i$, is upper bounded by} 
\begin{equation}\label{equ-Probability of unit vector as minimizer of f}
Pr\big( \mathcal{R}(\b{h}_L,\b{e}_i) \leq \mathcal{R}(\b{h}_L,\b{a}) \big) \leq e^{-LE(L)},
\end{equation}
\textit{where $\b{a}$ is any integer vector that is \textbf{not} a unit vector and  $E(L)=(1-\frac{3}{L})\log{\|\b{a}\|}$.}
\end{theorem}
As a consequence, when the number of users is large, there is only a single user which the relay is interested in decoding. Thus, all other users are considered as noise, and due to the decoding process of CF, each one contributes "self" noise, expressed by the approximation error of its channel gain to zero. This "self" noise degrades the achievable rate significantly, which eventually goes to zero as $L$ grows. Therefore, a restriction on the number of simultaneously transmitting users must be made in order to have a rate which does not go to zero. Thus, in this work, we examine such user scheduling in the context of CF schemes, devise an asymptotically optimal algorithm, and show that in this case the complex optimization on $\b{a}$ itself is straightforward, resulting in a very efficient algorithm.

%%%%%%%%%%%%%%%%%%%%%%%%%%%%%%%%%%%%%%%%%%%%%%%%%%%%%%%%%%%%%%%%%%%%%%%%%%%%%%%%%%%%%%

\section{Scheduling in CF}\label{sec-Scheduling in CF}

In this section, we present the specific scheduling problem of CF. Specifically, we do not describe the scheduling process itself, but rather center our interest on how the optimal schedule should be. We assume that in each transmission a subset of $k$ users are chosen by the scheduler. The total number of subsets is ${L \choose k}$, each having a channel vector which we denote by $\b{h}$, and a corresponding optimal vector $\b{a}$. 
\begin{definition}[The optimal schedule]
The optimal schedule is a subset of $k$ users which yields the highest sum of computation rates. The sum-rate achieved by this schedule is
\begin{equation}
\begin{aligned}
&\argmax_{\b{h} \in \mathcal{H^S},\b{a}}\left\{\sum_{i=1}^k \mathds{1}_{\{a_i \neq 0\}}\cR(\b{h},\b{a})\right\},				
\end{aligned}
\end{equation}
where $\mathcal{H^S}$ is the set of all vectors of length $k$ out of the channel vector $\b{h}_L$. 
\end{definition}

Note that maximizing the sum-rate of a single transmission may not suffice to achieve this rate in the long-run, as one has to make sure these linear combinations indeed sum up to a full rank matrix. We will discusse these consideration in the sequel.

The scheduling problem consists of two highly connected optimization problems. The first can be viewed as finding the proper $\b{h}$, and the second is finding the proper $\b{a}$ (for that $\b{h}$). A naive solution for this scheduling problem is to compute the sum-rate for all subsets of size $k$ (searching over all $\b{h}$ and the matching $\b{a}$) and choose the maximum among them. Since one has ${L \choose k} \approx L^k$ subsets, and for each subset there are $O(k^2\sqrt{1+P\|\b{h}^2\|})$ candidate coefficient vectors, the complexity is polynomial in $L$ but exponential in $k$. In fact, even for fixed $k$, such a complexity might be too high if $L$ is large.  In this work, we provide a polynomial time (in both $k$ and $L$) scheduling algorithm that finds the asymptotically (with $L$) optimal schedule for any fixed $P$.

\subsection{Scheduling algorithm}

The scheduling algorithm seeks a subset of users which has a channel vector which best fits a fixed coefficients vector, $\b{a^1} \triangleq (a_1,a_2,...,a_k)$ such that $|a_i|=1, \ \forall i$. By this choice, the coefficients vector has non-zero entries and has the smallest norm value (out of all vectors with all non-zero entries), i.e., $\sqn{a^1}=k$. Note that this definition defines a set of $2^k$ coefficients vectors, denoted as $\b{a^{\{1\}}}$, which corresponds to the possible differences in the signs of the elements. In this case, since there are no zero entries, the sum-rate will be the achievable rate of the scheduled subset times $k$. Thus, the algorithm seeks the schedule which maximizes the sum-rate: 
\begin{equation}\label{equ-the maximization problem of the scheduler}
	k \cdot \max_{\b{h} \in \mathcal{H^S}}\left\{\max_{\b{a} \in \b{a^{\{1\}}}}\left\{\cR(\b{h},\b{a})\right\}\right\}.
\end{equation}

\begin{algorithm}[t]
\caption{Optimal schedule for all 1 vector}
\label{algo-scheduling algorithm}
\hspace*{\algorithmicindent} \textbf{Input: $(\b{h}_L,\b{a})$} \\
 \hspace*{\algorithmicindent} \textbf{Output: $(\b{h}^*,\b{a}^*)$}
\begin{algorithmic}[1]
\Algphase{Initialization:}
\State $\b{h}_L^s \gets sort(Abs(\b{h}_L))$ 
\State $\b{h}_L^I \gets \text{ordering of } \b{h}_L^s \text{ in } \b{h}_L$
\State $\b{h}_L^{sign} \gets \b{h}_L^s./Abs(\b{h}_L^s)$ \Comment{element-wise devision}
\State $k \gets length(\b{a})$
\State $R^* \gets  0$
\State $i^* \gets -1$
\State $\b{h}^* \gets \emptyset$
\State $\b{a}^* \gets \emptyset$
\Algphase{Main:}
\For {$i =1; \ i\leq L-k; \ i++$}
	\State $\b{h} \gets  \b{h}_L^s(i:i+k)$
	\If {$\cR(\b{h},\b{a}) > R^*$}
		\State $i^* \gets i$
		\State $R^* \gets \cR(\b{h},\b{a})$
	\EndIf
\EndFor
\If {$i^* = -1$} 
	\Return $(\b{h}_L(1:k),\b{a})$
\Else
	\State $\b{h}^* \gets \b{h}_L(\b{h}_L^I(i^*):\b{h}_L^I(i^*)+k-1)$
	\State $\b{a}^* \gets \b{a}.*\b{h}_L^{sign}(i^*:i^*+k-1)$ 
\EndIf\\
\Return $(\b{h}^*,\b{a}^*)$
\end{algorithmic}
\end{algorithm}

The following Lemma shows an important property of the optimal coefficients vector which maximizes the achievable rate.
\begin{lemma}\label{lem-the signs of the coefficients vector are ruled by the channel}
The optimal vector $\b{a}$ satisfies either, $sign(h_i)=sign(a_i)$ for all $i$ or  $sign(h_i) \neq sign(a_i)$ for all $i$.
\end{lemma}
\begin{IEEEproof}
Considering the rate expression \eqref{equ-Computation rate with MMSE}, since $\sqn{a}$ does not depend on the signs, the optimal $\b{a}$ must maximize the inner product $(\b{h}^T\b{a})^2$. Obviously, all signs must match in order to have only positive elements in the summation of the inner product. 
\end{IEEEproof}

Lemma \ref{lem-the signs of the coefficients vector are ruled by the channel} implies that the inner maximization in \eqref{equ-the maximization problem of the scheduler} is trivial, since given a subset of channel coefficients $\b{h} \in \mathcal{H^S}$, the optimal $\b{a} \in \b{a^{\{1\}}}$ is clear - just set the signs according to those of $\b{h}$. Consequently, the following procedure is optimal for solving \eqref{equ-the maximization problem of the scheduler}: disregard the signs in $\b{h}_L$; find the optimal subset $(|h_1|,|h_2|,...,|h_k|)$, a one which best fits $\b{a}=(1,1,...,1)\triangleq \b{1}$; then simply set the signs of $\b{a}$ from all positive to the original signs of $\b{h}$. This reduces the double optimization in \eqref{equ-the maximization problem of the scheduler}, with $2^k$ options in the inner one, to a much simpler optimization:
\begin{equation}
	k\max_{\b{h} \in \mathcal{H^S}}\left\{\cR(|\b{h}|,\b{1})\right\}.
\end{equation}

The following lemma shows that for the case of all-ones coefficients vector, this search can be simplified after sorting the channel vector $\b{h}_L$ by the elements' absolute value. Thus, let us define $\b{h}_L^s$ as $|\b{h}_L|$ ordered in an ascending order.

\begin{lemma}\label{lem-optimal schedule for all one vector}
The optimal subset for the all-ones vector $\b{1}$ is a subset of $k$ consecutive elements in $\b{h}_L^s$. That is,
\begin{equation*}
\max_{\b{h} \in \mathcal{H^S}}\left\{\cR(|\b{h}|,\b{1})\right\}=\max_{i}\left\{\cR(|\b{h}_i'|,\b{1})\right\},
\end{equation*}
where $\b{h}_i'= (h_{L,i}^s,...,h_{L,i+k-1}^s)$ for $i \in [1,...,L-k+1]$.
\end{lemma}
\begin{IEEEproof}
In order to show this property we refer to another expression for the achievable rate {\cite[Theorem 1]{nazer2011compute}},
\begin{equation}
\cR(\b{h},\b{a})= \max \limits_{\alpha \in \R} \frac{1}{2} \log^+ \left( \frac{P}{\alpha^2+P\|\alpha \b{h}-\b{a}\|^2} \right).
\end{equation}
Note that if $\alpha=\frac{P\b{h}^T\b{a}}{1+P\|\b{h}\|^2}$, i.e., the MMSE coefficient which maximizes the rate, we obtain the rate expression as presented in equation \eqref{equ-Computation rate with MMSE}.
However, for a general and fixed $\alpha$ we have,
\begin{equation}
\begin{aligned}
&\max_{\b{h} \in \mathcal{H^S}}\left\{\cR(|\b{h}|,\b{1})\right\}\\
&=\max_{\b{h} \in \mathcal{H^S}}\left\{ \frac{1}{2} \log^+ \left( \frac{P}{\min \limits_{\alpha \in \R} \left\{\alpha^2+P\|\alpha |\b{h}|-\b{1}\|^2\right\}} \right)\right\}\\
&= \frac{1}{2} \log^+ \left( \frac{P}{\min \limits_{\b{h} \in \mathcal{H^S}} \left\{\min \limits_{\alpha \in \R} \left\{\alpha^2+P\|\alpha |\b{h}|-\b{1}\|^2\right\}\right\}} \right)\\
&= \frac{1}{2} \log^+ \left( \frac{P}{\min \limits_{\alpha >0} \left\{\min \limits_{\b{h} \in \mathcal{H^S}} \left\{\alpha^2+P\|\alpha |\b{h}|-\b{1}\|^2\right\}\right\}} \right),\\
\end{aligned}
\end{equation}
where in the last line we can reduce the minimization to $\alpha >0$ since for $\alpha <0$ we would increase the term for all $|\b{h}|$. Therefore we need to show that for any $\alpha >0$ 
\begin{equation*}
\argmin \limits_{\b{h} \in \mathcal{H^S}} \left\{\|\alpha |\b{h}|-\b{1}\|^2\right\}=\b{h}_i',
\end{equation*}
for  some $i \in [1,...,L-k+1]$.

Let us define the sequence $\Delta_j=(\alpha h_{L,j}^s-1)^2$, for $j=1,...,L$. This sequence can be monotonic increasing, monotonic decreasing or monotonic decreasing and then monotonic increasing with $j$; it depends on the value of $\alpha h_{L,1}^s$ and $\alpha h_{L,L}^s$ with respect to $1$. For example, if $\alpha h_{L,1}^s>1$ then the sequence is monotonic increasing with $j$.
Let us choose some $\b{h} \in \mathcal{H^S}$ such that its corresponding elements in $\b{h}_L^s$ are not consecutive. Hence, w.l.o.g. assume that two elements in $\b{h}$ corresponds to two elements $h_{L,i}^s$ and $h_{L,j}^s$ such that $i+1 \neq j $. Accordingly, either the choices $\b{h}_j'$ or $\b{h}_{j-k}'$ will minimize  $\left\{\|\alpha |\b{h}|-\b{1}\|^2\right\}$ since in at least one of the choice we would decrease with the sequence $\Delta_i$. Note also that this is true for the choices $\b{h}_i'$ or $\b{h}_{i-k}'$
\end{IEEEproof}

Considering Lemmas \ref{lem-the signs of the coefficients vector are ruled by the channel} and \ref{lem-optimal schedule for all one vector}, the optimal algorithm for the optimization problem in \eqref{equ-the maximization problem of the scheduler} is presented as Algorithm \ref{algo-scheduling algorithm}. Accordingly, the complexity of the algorithm is $O((L-k)L\log{L})$ due to the sorting of $\b{h}_L$ and the scan of $L-k$ scheduling options. The performance are summarized in the following theorem, whose proof is given in the sequel.

\begin{theorem}\label{the-optimality of algorithm 1}
\textit{Algorithm \ref{algo-scheduling algorithm} attains the optimal scaling laws of the expected sum-rate, which is $O(\log\log{L})$.}
\end{theorem}
Theorem \ref{the-optimality of algorithm 1} implies that the choice of fixing the coefficients vector $\b{a} \in \b{a^{\{1\}}}$ is asymptotically optimal as $L$ grows.

\begin{figure}[t]
    \centering
        \includegraphics[width=0.5\textwidth]{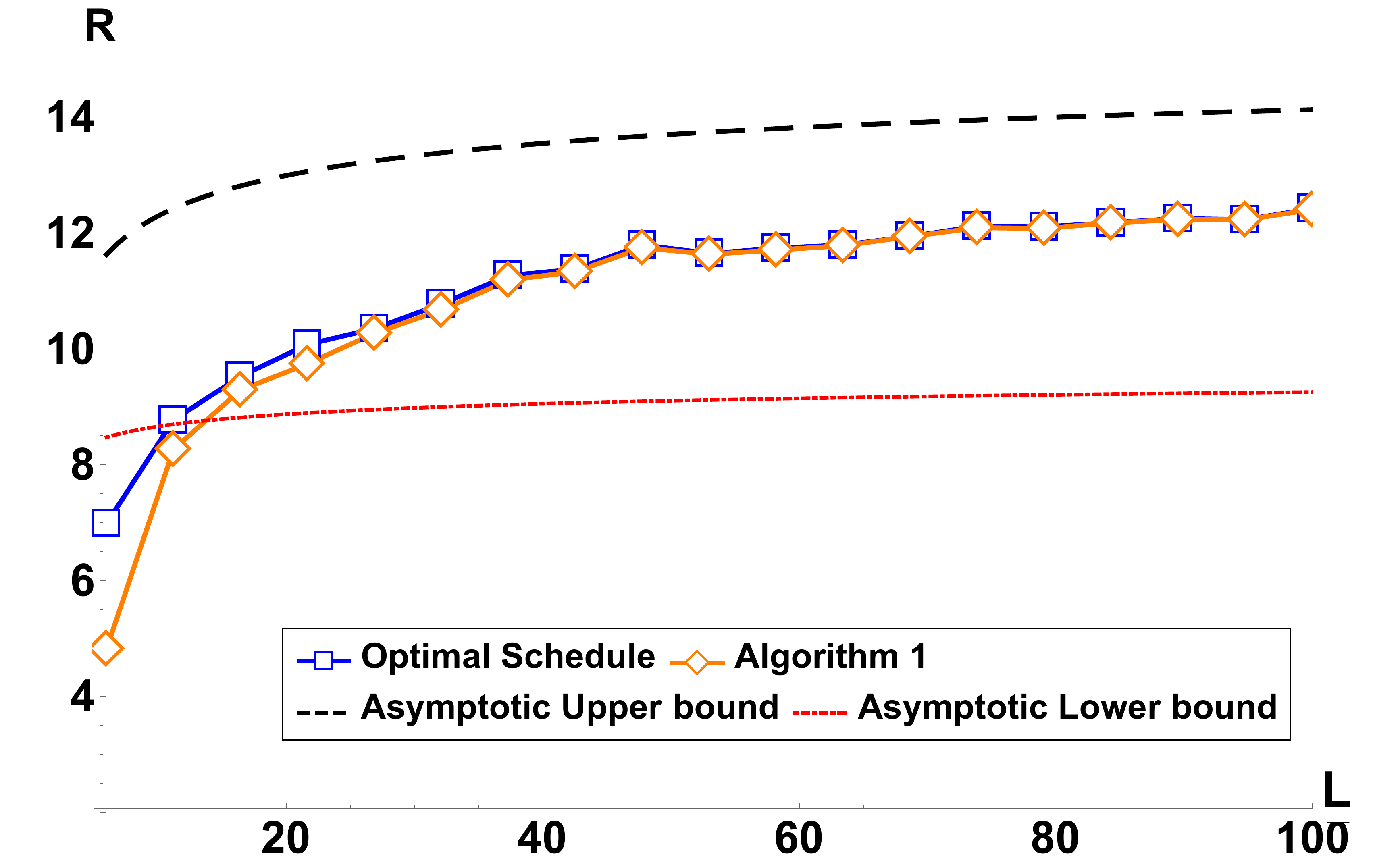}
    \caption{The system's sum-rate of the scheduling algorithm compared with the optimal scheduled sum-rate for $k=3$, as a function of $L$ with $P=100$. The asymptotic lower and upper bound which was given in Theorems \ref{the-Expected sum-rate of scheduling algorithm lower bound} and \ref{the-Expected sum-rate of scheduling algorithm upper bound}, respectively, are also plotted. The lower bound was plotted with $\delta=0.005$.}
       \label{fig-Optimal schedule CF comparisons}
\end{figure}

Simulation results of the system's expected sum-rate for Algorithm \ref{algo-scheduling algorithm}, compared with the optimal schedule (the naive solution) for $k=3$ as a function of $L$ are depicted Figure \ref{fig-Optimal schedule CF comparisons}. This simulation was compared also with the asymptotic upper and lower bounds in Theorems \ref{the-Expected sum-rate of scheduling algorithm lower bound} and \ref{the-Expected sum-rate of scheduling algorithm upper bound} below, which show good agreement even for moderate values of $L$. 

%%%%%%%%%%%%%%%%%%%%%%%%%%%%%%%%%%%%%%%%%%%%%%%%%%%%%%%%%%%%%%%%%%%%%%%%%%%%%%%%%%%%%%

\section{Sum-Rate Behaviour and the Scaling Law}
The proposed algorithm promises to find the optimal subset of users which attains the maximum sum-rate while fixing the coefficients vector $\b{a}$ such that $\b{a} \in \b{a^{\{1\}}}$. In this section, we start with a graphical interpretation for the problem of finding the optimal schedule; then, we provide a lower bound for Algorithm \ref{algo-scheduling algorithm} and compare it to a global upper bound on the achievable rate \cite{nazer2011compute}. Using it, we conclude that the scaling law of the suggested algorithm is similar to the best performance any scheduled subset can achieve with CF, giving Theorem \ref{the-optimality of algorithm 1}.

\subsection{Achievable rate under scheduling}
We now provide a graphical interpretation for the problem of finding the optimal schedule. This interpretation is based on the analysis of an upper bound on the achievable rate, yet gives the motivation for the suggested algorithm. Specifically, it explains the reason for ignoring scheduling opportunities which attain insignificant improvement in the achievable rate.

\begin{figure*}[t!]
    \centering
    \begin{subfigure}[b]{0.45\textwidth}
        \includegraphics[width=\textwidth]{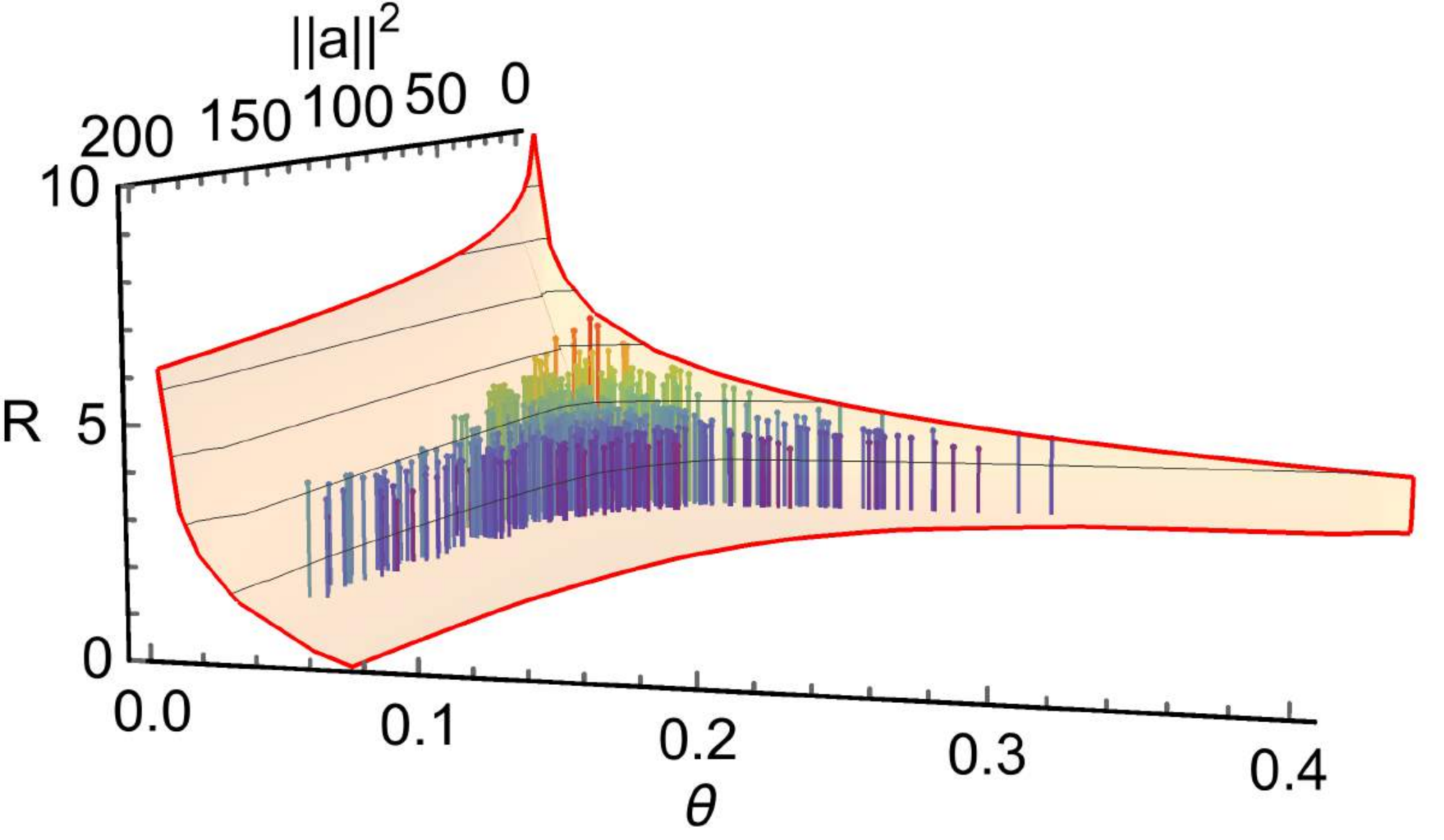}
        \caption{$L=15$}
        \label{fig-Rate_Large_P_func_theta_norm_combined_L=15}
    \end{subfigure}
    \begin{subfigure}[b]{0.45\textwidth}
        \includegraphics[width=\textwidth]{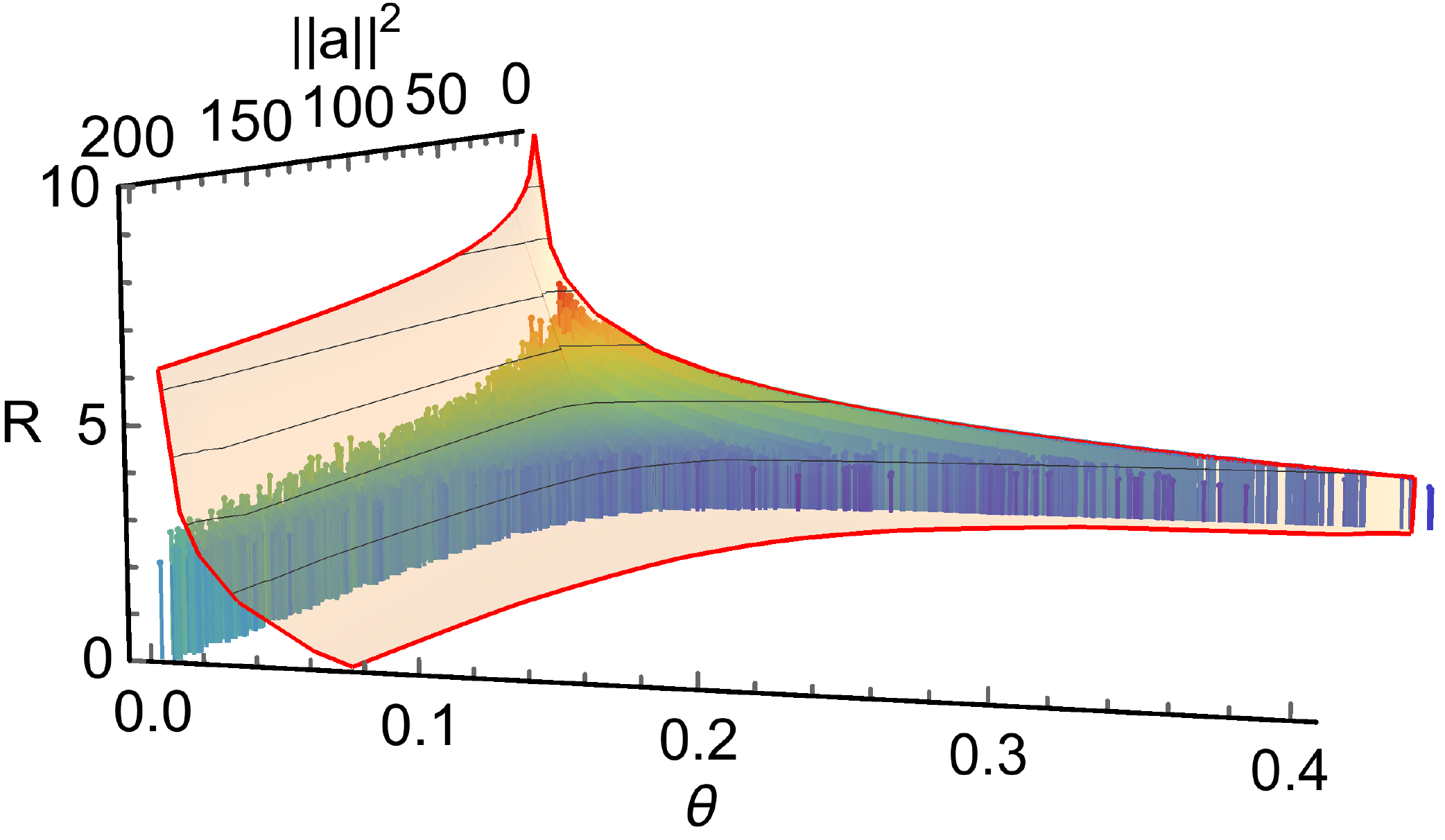}
        \caption{$L=45$}
        \label{fig-Rate_Large_P_func_theta_norm_combined_L=45}
    \end{subfigure}
    \caption{The achievable rate as a function of $\theta$, i.e., the angle between $\b{a}$ and $\b{h}$, and the squared norm of $\b{a}$. The discrete points are simulation results for the rate of each subset of users of size $k=3$ for a specific realization of the channel vector $\b{h}_L$ with $(a)L=15$, $(b)=L=45$ and $P=1000$. The transparent curved plane describes an upper bound on the rate as in \eqref{equ-Achievable rate high SNR}, where, for ease of visualization, a continues function was plotted.}
    \label{fig-Rate_Large_P_func_theta_norm_combined}
\end{figure*}

Consider the achievable rate of a given schedule:

\begin{equation}\label{equ-Achievable rate high SNR}
\begin{aligned}
\cR(\b{h},\b{a})&= \frac{1}{2} \log^+ \left( \left( \|\b{a}\|^2- \frac{P(\b{h}^T\b{a})^2}{1+P\|\b{h}\|^2} \right)^{-1}\right)\\
		      &\underset{P \rightarrow \infty}{\longrightarrow}\frac{1}{2} \log^+ \left( \left( \|\b{a}\|^2- \frac{(\b{h}^T\b{a})^2}{\|\b{h}\|^2} \right)^{-1}\right)\\	
      		      &=\frac{1}{2} \log^+ \left( \left( \|\b{a}\|^2- \|\b{a}\|^2\cos^2(\theta) \right)^{-1}\right)\\			
      		      &=\frac{1}{2} \log^+ \left(\left(  \|\b{a}\|^2\sin^2(\theta) \right)^{-1}\right),\\			
\end{aligned}
\end{equation}
where $\theta$ is the angle between $\b{h}$ and its coefficients vector $\b{a}$.

Figure \ref{fig-Rate_Large_P_func_theta_norm_combined} depicts the behaviour of the achievable rate as a function of $\theta$ and $\sqn{a}$. The discrete lines represent simulation results for the achievable rate of each subset of size $k=3$, out of  a specific realization of the channel vector $\b{h}_L$. $P=1000$. The continuous curve is a  graphic representation of equation \eqref{equ-Achievable rate high SNR}. 

The continuous curve is a bit misleading since, for one, $\b{a}$ is an integer vector, hence, its squared norms takes only integer values. Second, for a certain $\|\b{a}\|^2$ there are a finite possible choices of $\b{a}$, e.g., for $\|\b{a}\|^2=5$ and dimension 2 the possible vectors are only $(1,2),\ (-1,2),\ (1,-2)$ and $(-1,-2)$. That is, in this case there are 4 possible angles for a given $\b{h}$.  Thus, this curve should look like a discrete plot. Yet, for ease of visualization and to recognize the rate behavior, we plotted a continuous curve. Note that the curve consider all integer vectors and not only the optimal for a certain $\b{h}$. 

There are several observations which can be inferred from Figure \ref{fig-Rate_Large_P_func_theta_norm_combined}. The slope of the rate as a function of $\theta$ is much sharper than the slope of the rate as a function of $\sqn{a}$, with an exception for the smallest values of $\sqn{a}$ where a tip is noticeable. For the minimum possible value of $\sqn{a}$, i.e. a unit vector, the rate is non zero for all angles in particular for small values of angels. This suggests that the rate is far more sensitive to small changes in the angle than small changes in $\sqn{a}$. That is, for a given $\b{h}$, if the angle between the coefficients vector and $\b{h}$ is small, a high norm may be sustainable without much loss in optimality. On the other hand the opposite is not true.

Another important observation is that, as the values of $\sqn{a}$ grows, the slope is very small and thus we may only need to seek the scheduling solution in the dimension of $\theta$ without the risk of significant rate loss. In general we would still prefer $\sqn{a}$ with low norm due to its penalty on the rate.

We emphasis that the assumption of $P \rightarrow \infty$ essentially means that the search domain becomes infinite. That is, for each given $\b{h}$ we can find an excellent collinear integer approximation by considering vectors with very high norm in order to decrease the angle between the vectors. In this case, one should also consider the complexity of this search which is polynomial with $P$. 

Figure \ref{fig-Rate_Large_P_func_theta_norm_combined} can be explained also in the following manner. In the rate maximization problem, for a certain $\b{h}$, we have a sample (of points) out of the continuous curve in Figure \ref{fig-Rate_Large_P_func_theta_norm_combined} (since not all values of the angles are possible). The optimal rate is a single point out of this sample. While in the scheduling problem, each schedule is a different sample and the scheduler chooses the optimal point out of all the samples. Naturally, the optimal points for all possible schedules will be placed as close as possible to the boundaries and close to zero on the angle axis where the rate is high. One can imagine it as follows. For a given dimension and a fixed upper bound on the norm value we have a finite collection of possible $\b{a}$ vectors with a certain $\sqn{a}$ value. That is, as the number of $\b{h}$ vectors grows, i.e. $L$ grows, we get a reacher plot, i.e., more points will be added to the graph as can be seen from figure \ref{fig-Rate_Large_P_func_theta_norm_combined_L=15} and \ref{fig-Rate_Large_P_func_theta_norm_combined_L=45} respectively.

It is clear from Figure \ref{fig-Rate_Large_P_func_theta_norm_combined} that the highest rates are obtained for small norm values. Specifically, as $L$ grows ($k$ is still fixed) the optimal coefficients vector that attains the highest achievable rate is the unit vector. This can also be explained as a consequence of \cite[Theorem 1]{shmuel2017necessity}, which shows the superiority, in probability, of a unit vector with comparison to a certain non-trivial coefficients vector. For example, for $k=3$, the probability for a certain non-trivial vector to be chosen as the optimal one, comparing to a unit vector, is at most $0.3$. However, other low norm vectors attain significant high rate as well. This is a significant factor in terms of the sum-rate: a specific schedule may have low-norm (contains zeroes) coefficients vector, which gives high achievable rate, but its sum-rate may be small with respect to other vectors which have more non-zero entries. 

\begin{figure}[t!]
\center
        \includegraphics[width=0.35\textwidth]{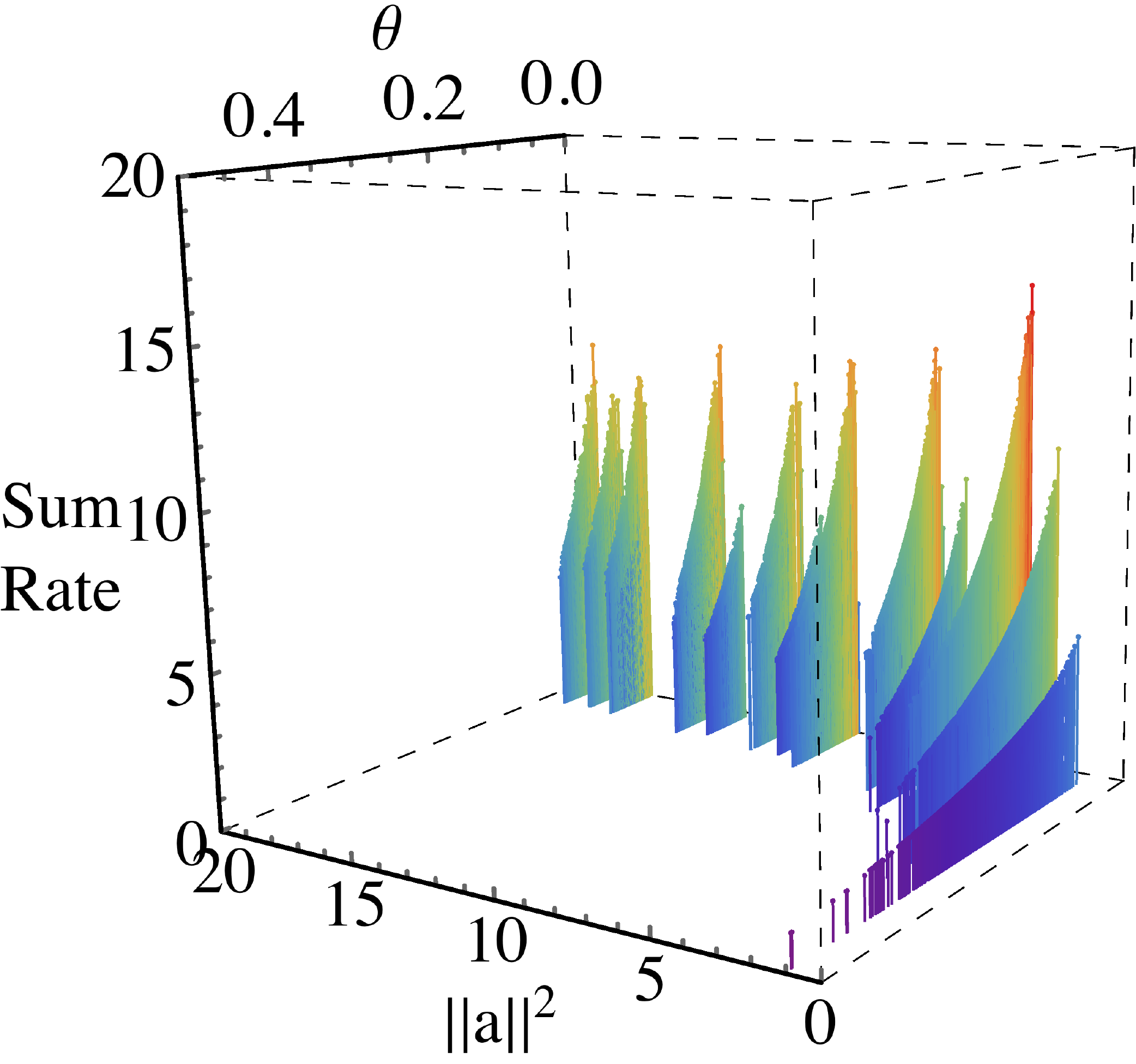}
    \caption{The sum-rate of each subset of users as a function of the squared norm of the optimal $\b{a}$ for the subset's channel vector and as a function of the angle between these vectors. Where $P=1000$, $L=45$ and $k=3$ for different number of users.}
    \label{fig-SumRate_Large_P_func_theta_norm_a_OPT}
\end{figure}

Thus, in terms of the sum-rate, it may be beneficial to schedule groups which have no zero entries in their coefficients vectors (with relatively high achievable rate). Figure \ref{fig-SumRate_Large_P_func_theta_norm_a_OPT} depicts simulation results of the \emph{sum-rate} of each subset of users of size $k=3$, for a specific realization of the channel vector $\b{h}_L$. The right most hand curve corresponds to $\sqn{a}=1$, i.e., a unit vector gives very low sum-rate due to the presence of the $k-1$ zeros. On the other hand, one can notice that the highest sum-rates are obtained for coefficients vector with $\sqn{a}=3$, which is the smallest norm value with no zeros. 

Consequently, in order to find the optimal schedule, we expect to use only a small set of fixed coefficients vectors, which have a small norm, with no zero entries as a good schedule. Thus, when searching for the optimal schedule, we flip the order in our optimization: we fix a reasonably good $\b{a}$, and search for the best $\b{h}$. As it turns out, this will be asymptotically optimal.

\subsection{Best channel for a fixed $\b{a}$}\label{subsec-Best channel for a fixed a}

The polynomial algorithm as given in \cite{sahraei2014compute} finds the optimal coefficients vector $\b{a}$ for a given channel vector $\b{h}$. We now consider the opposite case in which we fix a specific $\b{a}$ and seek the optimal $\b{h}$ (that is, the optimal subset of senders) which maximize the achievable rate. We have, 

\begin{equation}\label{equ-Scheduling optimal rate fixed a}
\begin{aligned}
&\argmax_{\b{h} \in \mathcal{H^S}}\left\{\cR(\b{h},\b{a})\right\} \\
				&= \argmax_{\b{h} \in \mathcal{H^S}}\left\{\frac{1}{2} \log^+ \left( \|\b{a}\|^2- \frac{P(\b{h}^T\b{a})^2}{1+P\|\b{h}\|^2} \right)^{-1}\right\}\\
				&=\frac{1}{2} \log^+ \left( \|\b{a}\|^2-  \argmax_{\b{h} \in \mathcal{H^S}}\left\{\frac{P(\b{h}^T\b{a})^2}{1+P\|\b{h}\|^2}\right\} \right)^{-1}\\
				&=\argmax_{\b{h} \in \mathcal{H^S}}\left\{\frac{P(\b{h}^T\b{a})^2}{1+P\|\b{h}\|^2}\right\}\\
				&=\argmax_{\b{h} \in \mathcal{H^S}}\left\{\frac{P\|\b{h}\|^2\|\b{a}\|^2\cos^2(\theta)}{1+P\|\b{h}\|^2}\right\}\\
				&=\argmax_{\b{h} \in \mathcal{H^S}}\left\{\frac{\cos^2(\theta)}{1+\frac{1}{P\|\b{h}\|^2}}\right\}.				
\end{aligned}
\end{equation}
Thus, the $\b{h}$ which maximizes the achievable rate has a small angle with $\b{a}$ and a high norm. However, this causes a tradeoff, since the highest norm vector may not be the one with the smallest angle to $\b{a}$. The scheduler should seek the optimal tradeoff point to maximize the achievable rate. 

\begin{figure*}[t]
    \centering
    \begin{subfigure}[b]{0.3\textwidth}
        \includegraphics[width=0.95\textwidth]{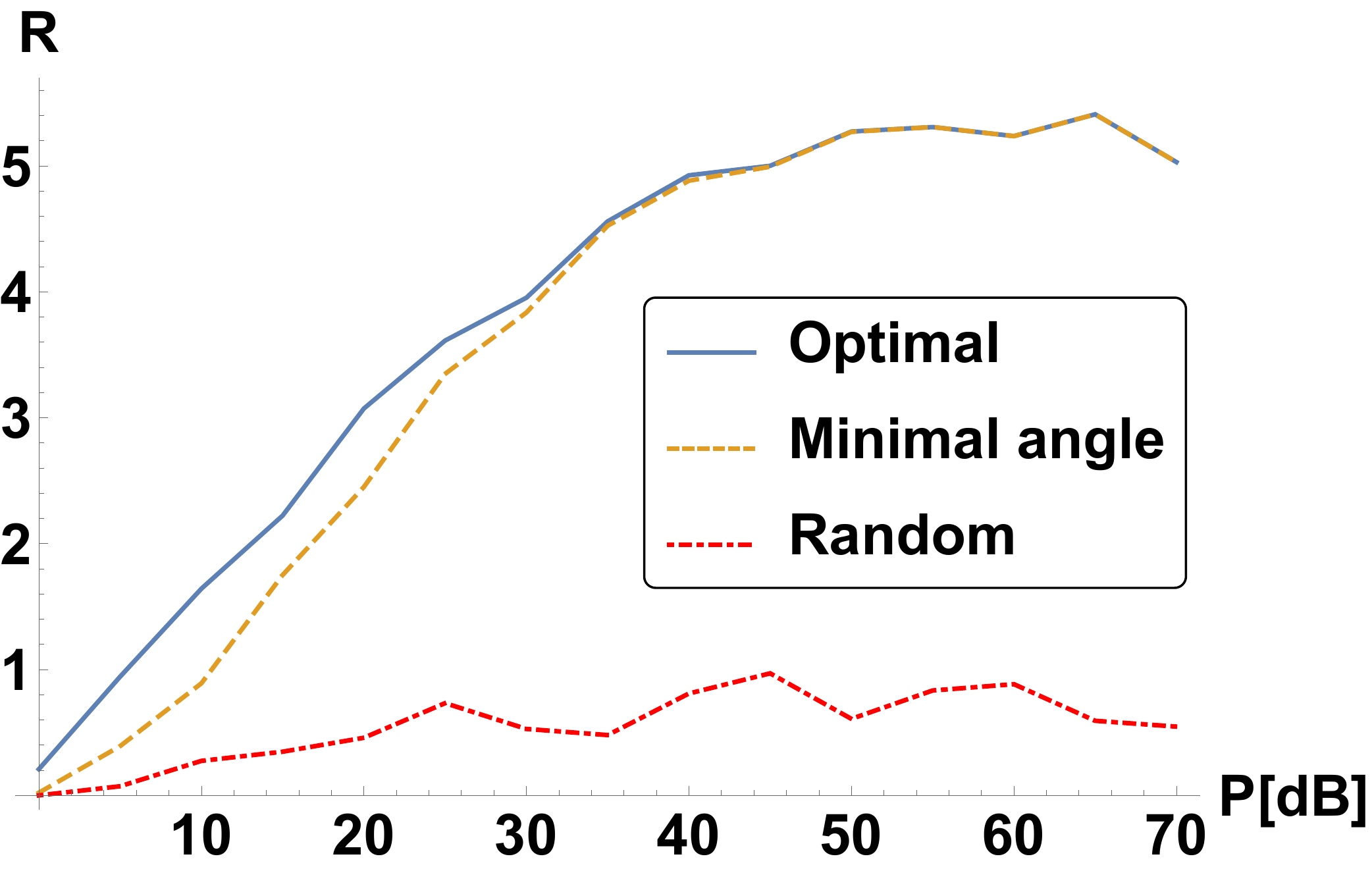}
        \caption{$\b{a}=(2,1,1)$}
    \end{subfigure}
    \begin{subfigure}[b]{0.3\textwidth}
        \includegraphics[width=0.95\textwidth]{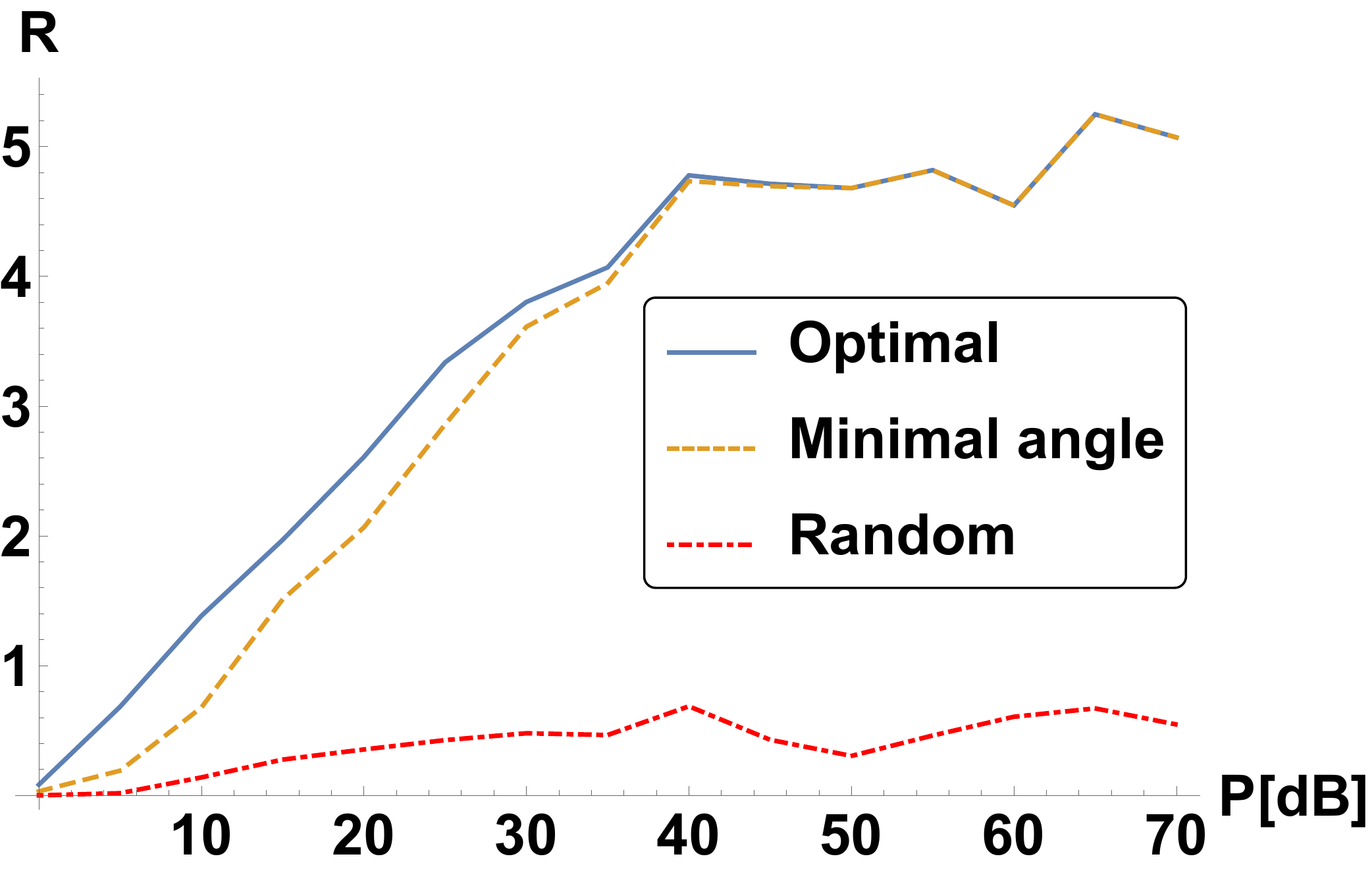}
        \caption{$\b{a}=(2,2,1)$}
    \end{subfigure}
    \begin{subfigure}[b]{0.3\textwidth}
        \includegraphics[width=0.95\textwidth]{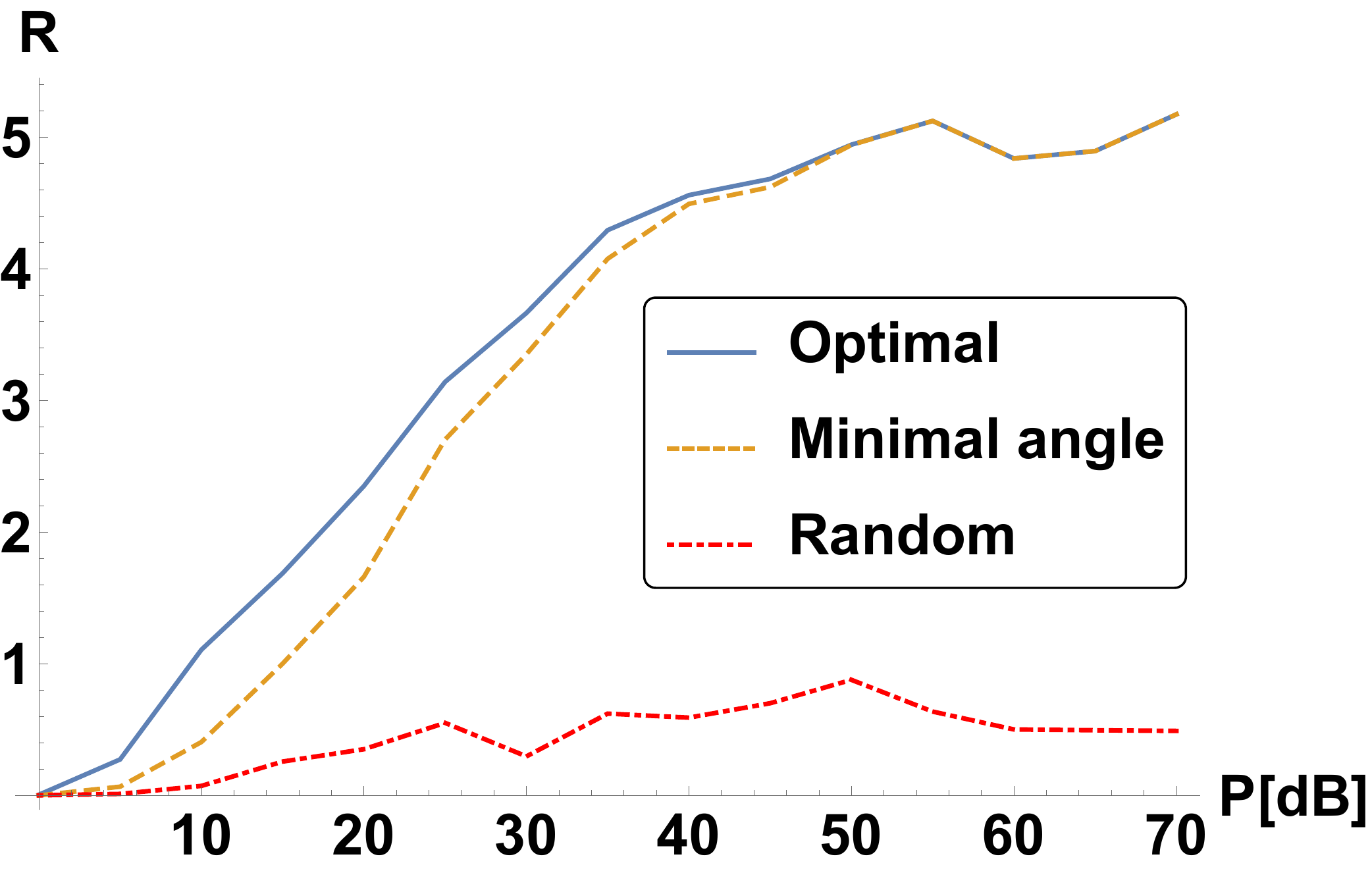}
        \caption{$\b{a}=(3,2,1)$}
    \end{subfigure}
    \caption{Achievable rate for scheduling $k=3$ out of $L=20$ users with different fixed coefficients vectors. The users were chosen randomly (dot-dashed), optimally (solid), or according to the minimal angle between the channel vector of the scheduled group and $\b{a}$.}
    \label{fig-Scheduling_angle_vs_opt_funcOf_P_fixed_a}
\end{figure*}

Considering the rate expression \eqref{equ-Scheduling optimal rate fixed a} for the regime of high SNR, i.e. $P \rightarrow \infty$, where we are left only with

\begin{equation}\label{equ-Scheduling optimal rate fixed a high SNR}
\argmax_{\b{h} \in \mathcal{H^S}}\left\{\cos^2(\theta)\right\},				
\end{equation}
which essentially mean that the scheduler should only seek the group which has the smallest angle to $\b{a}$ as the optimal scheduling policy. This can be seen in Figure \ref{fig-Scheduling_angle_vs_opt_funcOf_P_fixed_a} where we fixed various coefficients vectors and plotted the achievable rate for choosing the channel vector with the smallest angle comparing to the optimal choice (i.e. the schedule which gives the maximal rate) and a random choice. We note here that, for the case of $P \rightarrow \infty$, the scheduler eventually encounter the problem of choosing the maximum out of $L \choose k$ r.vs. which are distributed as $Beta(\frac{1}{2},\frac{k-1}{2})$. We also note that some of these r.vs. are dependent due to the fact that $\mathcal{H^S}$ is the set of all possible sub-sets of $\b{h}_L$ which make it hard to analyse.

\subsection{Asymptotic guarantees}
We now give a lower bound on the performance of Algorithm \ref{algo-scheduling algorithm}. In particular, we show that asymptotically with $L$, the choice of an all-1 coefficients vector is optimal. 
\begin{theorem}\label{the-Expected sum-rate of scheduling algorithm lower bound}
\textit{The expected sum-rate of Algorithm \ref{algo-scheduling algorithm} is lower bounded by the following,}
\begin{multline*}
	\EX\left[k\max_{\b{h} \in \mathcal{H^S}}\left\{\cR(\b{|h|},\b{1})\right\}\right] \geq \\ \frac{k}{2}  \log^+ \left( k \left(1-\frac{Pku^4}{(u+\delta)^2(1+Pku^2)}(1-o(1)) \right)\right)^{-1},
\end{multline*}
\textit{where $u=\sqrt{2\ln{2\sqrt{L}}}-\delta$ and $\delta$ is some small constant greater than zero.}

\textit{Thus, the expected sum-rate for Algorithm \ref{algo-scheduling algorithm} scales at least as $O(\frac{k}{4}\log{\log{L}})$. $o(1) \rightarrow 0$ as $L \rightarrow \infty$.}
\end{theorem}

The values $u$ and $\delta$ were chosen such that, with probability that goes to one with $L$, there are at least $k$ users with channel fading coefficients in the range $[u,u+\delta]$. Thus, we can lower bound the magnitude of the channel coefficients of the scheduled subset using $u$, and upper bound the angle between $\b{h}$ and $\b{1}$ using $\delta$. This bound applies (asymptotically with $L$) on the performance of Algorithm \ref{algo-scheduling algorithm} since the best $k$ out of these users will be chosen. Theorem \ref{the-Expected sum-rate of scheduling algorithm lower bound} indicates that indeed, as the number of users grows, the system's sum-rate grows as well, making scheduling not only mandatory but worthwhile. The proof tor Theorem \ref{the-Expected sum-rate of scheduling algorithm lower bound} is given below.

In \cite{nazer2011compute}, the following universal upper bound for the achievable rate was given
\begin{equation}\label{equ-universal upper bound on the achievable rate}
 	\cR(\b{h},\b{a}^{opt})\leq \frac{1}{2}\log{(1+P\max_i\{h_i^2\})},
\end{equation}
where $\b{h}$ is any channel vector of dimension $k$ and $\b{a}^{opt}$ is the coefficients vector which maximize the achievable rate.
Using this result, one can derive an upper bound on the expected performance of any scheduling algorithm and its scaling laws, at the limit of large $L$. 

\begin{theorem}\label{the-Expected sum-rate of scheduling algorithm upper bound}
\textit{The expected sum-rate of any scheduling algorithm, is upper bounded by the following,}
\begin{multline*}
	\EX\left[k\max_{\b{h} \in \mathcal{H^S}}\left\{\cR(\b{h},\b{a}^{opt})\right\}\right] \leq \\ \frac{k}{2}\log{\left(1+P\left(2\ln{L}-\ln{\ln{L}}-2\ln{\Gamma\left(\frac{1}{2}\right)}+\frac{\gamma}{2}+o(1)\right)\right)},
\end{multline*}
\textit{where $\gamma$ is the Euler-Mascheroni constant.}

\textit{Thus, the expected sum-rate for the suggested scheduling algorithm scales at most as $\frac{k}{2}\log{\log{L}}$. $o(1) \rightarrow 0$ as $L \rightarrow \infty$.}
\end{theorem}

Theorems \ref{the-Expected sum-rate of scheduling algorithm lower bound} and \ref{the-Expected sum-rate of scheduling algorithm upper bound} show that the Algorithm \ref{algo-scheduling algorithm} is asymptotically optimal as the upper and lower bounds on the performance scale as $O(\log{\log{L}})$. This proves Theorem \ref{the-optimality of algorithm 1}.

%%%%%%%%%%%%%%Proofs%%%%%%%%%%%%%%%
\begin{IEEEproof}[Proof of Theorem \ref{the-Expected sum-rate of scheduling algorithm lower bound}]
We have,
\small
\begin{equation}
\begin{aligned}\label{equ-expected sum-rate}
	&\EX\left[k\max_{\b{h} \in \mathcal{H^S}}\left\{\cR(\b{|h|},\b{1})\right\}\right] \\
	&=\EX\left[k\max_{\b{h} \in \mathcal{H^S}}\left\{ \frac{1}{2} \log^+ \left( k- \frac{P\left(\b{|h|}^T\b{1}\right)^2}{1+P\|\b{h}\|^2} \right)^{-1} \right\}\right] \\
	&=\EX\left[\frac{k}{2} \log^+ \left( k- \max_{\b{h} \in \mathcal{H^S}}\left\{ \frac{P\left(\b{|h|}^T\b{1}\right)^2}{1+P\|\b{h}\|^2} \right\} \right)^{-1}\right] \\
	&\overset{(a)}{\geq} \EX\left[\frac{k}{2} \log^+ \left( k-  \frac{P\left(\b{h'}^T\b{1}\right)^2}{1+P\|\b{h'}\|^2}  \right)^{-1}\right] \\
	&\overset{(b)}{\geq} \frac{k}{2} \log^+ \left( k- \EX\left[ \frac{P\left(\b{h'}^T\b{1}\right)^2}{1+P\|\b{h'}\|^2}\right]  \right)^{-1} \\
	&= \frac{k}{2} \log^+ \left( k- \EX\left[ \frac{Pk\sqn{h'}\cos^2(\theta')}{1+P\|\b{h'}\|^2} \right]  \right)^{-1}, \\
\end{aligned}
\end{equation}
\normalsize
where in $(a)$ we chose some specific $\b{h'} \in \mathcal{H^S}$ and $(b)$ follows from Jensen's inequality. As section \ref{subsec-Best channel for a fixed a} suggests, the optimal schedule should be a subset of users with a high norm channel vector and a small angle between its channel vector and the corresponding coefficients vector. Thus, let us define the values $u(L)$ and $\delta$ such that $\b{h'}$ maintains $u\leq | h_i' | \leq u+\delta, \ \forall i$. With this definition we are able to bound the parameters for a good schedule. With this definition, we are able to bound the parameters for a good schedule. The values of $u(L)$ and $\delta$ can help us tune the norm (by taking a high value of $u$) and the angle with $\b{1}$ (by taking a small value of $\delta$) to attain a better bound as a function of $L$. 
And let us define $P_r(\xi)$ as the probability of having at least $k$ elements in $\b{h}_L$ such that we can find an $\b{h'}$ satisfying the constraint above. We thus can write the last equation in \eqref{equ-expected sum-rate} as follows,

\begin{equation}\label{equ-the rate split to probabilities}
\small
\begin{aligned}
		&= \frac{k}{2} \log^+ \left( k- \left(\EX\left[ \frac{Pk\sqn{h'}\cos^2(\theta')}{1+P\|\b{h'}\|^2} \ \Big| \xi \right] P_r(\xi) + \right.\right. \\
		&\quad \quad \left.\left.\EX\left[ \frac{Pk\sqn{h'}\cos^2(\theta')}{1+P\|\b{h'}\|^2} \ \Big| \bar{\xi} \right] (1-P_r(\xi)) \right) \right)^{-1}\\	
		&\geq \frac{k}{2} \log^+ \left( k- \EX\left[ \frac{Pk\sqn{h'}\cos^2(\theta')}{1+P\|\b{h'}\|^2} \ \Big| \xi \right] P_r(\xi) \right)^{-1}.
		%&= \frac{k}{2} \log^+ \left( k- \EX\left[ \frac{Pk\sqn{h'}\cos^2(\theta')}{1+P\|\b{h'}\|^2} \ \Big| \xi \right]  \right)^{-1}P_r(\xi) + \\
		%&\quad \quad \frac{k}{2} \log^+ \left( k- \EX\left[ \frac{Pk\sqn{h'}\cos^2(\theta')}{1+P\|\b{h'}\|^2} \ \Big| \bar{\xi} \right]  \right)^{-1}(1-P_r(\xi))\\	
		%&\geq \frac{k}{2} \log^+ \left( k- \EX\left[ \frac{Pk\sqn{h'}\cos^2(\theta')}{1+P\|\b{h'}\|^2} \ \Big| \xi \right]  \right)^{-1}P_r(\xi).
\end{aligned}
\end{equation}
\normalsize
Considering the conditioning we can lower bound $\sqn{h'}$ and $\cos^2(\theta')$ as follows,
\begin{equation}\label{equ-lower bound on h' and cos}
\begin{aligned}
		&\sqn{h'} \geq ku^2;\\
		&\cos^2(\theta')=\frac{\left( \sum_{i=1}^{k} h_i' \right)^2}{k\sqn{h'}} \geq \frac{k^2u^2}{k^2(u+\delta)^2}=\frac{1}{1+\frac{2\delta}{u}+\frac{\delta^2}{u^2}}.
\end{aligned}
\end{equation}
The probability $P_r(\xi)$ can be computed using the binomial distribution with probability of success $p(u,\delta)=2(\Phi(u)-\Phi(u+\delta))$ where $\Phi$ is the CDF of the normal distribution and can be lower bounded using the Chernoff bound. That is,  

\begin{equation}\label{equ-lower bound on probability xi}
\begin{aligned}
P_r(\xi)&=\sum_{i=k}^{L} {L \choose i} p(u,\delta)^i(1-p(u,\delta))^{L-i}\\
	&=1-\sum_{i=0}^{k-1} {L \choose i} p(u,\delta)^i(1-p(u,\delta))^{L-i}\\
	&\geq 1-e^{-\frac{1}{2p(u,\delta)}\frac{(Lp(u,\delta)-(k-1))^2}{L}}.
\end{aligned}
\end{equation}
Note that in order that $P_r(\xi)$ will go to one with $L$, $p(u,\delta)$ must decay at most as $\frac{1}{\sqrt{L}}$. Therefore, we would like to find $u$ and $\delta$ which will maintain this behaviour. That is, we wish to find $u$ and $\delta$ such that,
\begin{equation}\label{equ-behaviour of p}
\begin{aligned}
	\lim_{L\rightarrow \infty} \frac{p(u,\delta)}{1/\sqrt{L}}= c,
\end{aligned}
\end{equation}
where $c \in (0,\infty]$. Thus,
\begin{equation}\label{equ-behaviour of p limit development}
\begin{aligned}
	&\lim_{L\rightarrow \infty} \frac{p(u,\delta)}{1/\sqrt{L}}\\
	&=\lim_{L\rightarrow \infty} \frac{2(\Phi(u)-\Phi(u+\delta))}{1/\sqrt{L}}\\
	&=\lim_{L\rightarrow \infty} \frac{\frac{1}{\sqrt{2\pi}}\int_{u}^{u+\delta}e^{-\frac{t^2}{2}}dt}{1/2\sqrt{L}}\\
	&\overset{(a)}{\geq} \lim_{L\rightarrow \infty} \frac{\delta\frac{1}{\sqrt{2\pi}}e^{-\frac{(u+\delta)^2}{2}}}{1/2\sqrt{L}}\\
	&\overset{(b)}{=} \lim_{L\rightarrow \infty} \frac{\delta\frac{1}{\sqrt{2\pi}}e^{-\frac{\sqrt{2\ln{(2\sqrt{L})}}^2}{2}}}{1/2\sqrt{L}}\\
	&= \lim_{L\rightarrow \infty} \frac{\delta}{\sqrt{2\pi}}
\end{aligned}
\end{equation}
Where in $(a)$ we bound the probability by the length of the interval and the density function smallest value in the interval $[u,u+\delta]$. Setting $u=\sqrt{2\ln{2\sqrt{L}}}-\delta$ in $(b)$ guarantees the desired outcome as long as $\delta$ is a constant grater than zero. Thus, for 

\begin{equation}\label{equ-probability p that will go to one}
\begin{aligned}
		p(u,\delta)&= \delta \frac{1}{\sqrt{2\pi}}e^{-\frac{u^2}{2}}\\ 
			&= \delta\frac{1}{\sqrt{2\pi}}e^{-\frac{\sqrt{2\ln{(2\sqrt{L})}}^2}{2}}\\
			&= \delta\frac{1}{\sqrt{2\pi}} \frac{1}{2\sqrt{L}},
\end{aligned}
\end{equation}
$P_r(\xi)$ will go to one with $L$. We note here that we require that $k-1<Lp(u,\delta)$ for the correctness of the Chernoff bound in \eqref{equ-lower bound on probability xi}. That is, $k<\frac{\delta}{2\sqrt{2\pi}} \sqrt{L}+1$.

Setting \eqref{equ-lower bound on h' and cos} and \eqref{equ-lower bound on probability xi} in \eqref{equ-the rate split to probabilities} we get 
\begin{equation}\label{equ-lower bound on the sum-rate final expression} 
\small
\begin{aligned}	
		&\frac{k}{2} \log^+ \left( k- \EX\left[ \frac{Pk\sqn{h'}\cos^2(\theta')}{1+P\|\b{h'}\|^2} \ \Big| \xi \right] P_r(\xi) \right)^{-1}\\
		&\geq \frac{k}{2}  \log^+ \left( k \left(1-\frac{1}{1+\frac{2\delta}{u}+\frac{\delta^2}{u^2}}\frac{1}{\frac{1}{Pku^2}+1} \right.\right.\\
		&\quad\quad \left.\left. \cdot\left(1-e^{-\frac{1}{2p(u,\delta)}\frac{(Lp(u,\delta)-(k-1))^2}{L}}\right) \right)\right)^{-1}\\
		&\overset{(a)}{=} \frac{k}{2}  \log^+ \left( k \left(1-\frac{1}{1+\frac{2\delta}{u}+\frac{\delta^2}{u^2}}\frac{1}{\frac{1}{Pku^2}+1}\left(1-o(1)\right) \right)\right)^{-1}\\
		&\overset{(a)}{=} \frac{k}{2}  \log^+ \left( k \left(1-\frac{Pku^4}{(u+\delta)^2(1+Pku^2)}\left(1-o(1)\right) \right)\right)^{-1},
\end{aligned}
\end{equation}
\normalsize
where $(a)$ follows from the asymptotic behaviour of the exponent when setting $p(u,\delta)$ as in \eqref{equ-probability p that will go to one}. This can be seen as follows,
\begin{equation}
\begin{aligned}
	&\lim_{L\rightarrow \infty} \frac{e^{-\frac{1}{2p(u,\delta)}\frac{(Lp(u,\delta)-(k-1))^2}{L}}}{1} \\
	&=\lim_{L\rightarrow \infty} e^{-\frac{1}{2\delta\frac{1}{\sqrt{2\pi}} \frac{1}{2\sqrt{L}}}\frac{\left(L\delta\frac{1}{\sqrt{2\pi}} \frac{1}{2\sqrt{L}}-(k-1)\right)^2}{L}}\\
	&=\lim_{L\rightarrow \infty} e^{-\frac{\sqrt{2\pi}\sqrt{L}}{\delta} {\frac{\left(  \frac{\sqrt{L}\delta}{2\sqrt{2\pi}}-(k-1)\right)^2}{L}}}\\
	&=\lim_{L\rightarrow \infty} e^{-\frac{\sqrt{2\pi}\sqrt{L}}{\delta}{\left(  \frac{\delta}{2\sqrt{2\pi}}-\frac{(k-1)}{\sqrt{L}}\right)^2}}\\
	&= e^{-\lim_{L\rightarrow \infty}\frac{\sqrt{2\pi}\sqrt{L}}{\delta}{\left(  \frac{\delta}{2\sqrt{2\pi}}-\frac{(k-1)}{\sqrt{L}}\right)^2}}\\
	&=e^{-\infty}=0.\\
\end{aligned}
\end{equation}
It can be verified (the computation mappears in Appendix \ref{Appendix A}) that the scaling laws of \eqref{equ-lower bound on the sum-rate final expression} indeed behave as $\frac{k}{4}\log{\log{L}}$ which completes the proof. 
\end{IEEEproof}

\begin{IEEEproof}[Proof of Theorem \ref{the-Expected sum-rate of scheduling algorithm upper bound}]
Since the universal bound in \eqref{equ-universal upper bound on the achievable rate} holds for all $\b{h}$, it holds for any subset of users as well. Thus,
\begin{align*}
	& \EX\left[k\max_{\b{h} \in \mathcal{H^S}}\left\{\cR(\b{h},\b{a}^{opt})\right\}\right]\\
	&\leq \EX\left[k\max_{\b{h} \in \mathcal{H^S}}\left\{ \frac{1}{2}\log{(1+P\max_i\{h_i^2\})} \right\}\right]\\
	&\overset{(a)}{=} \EX\left[\frac{k}{2}\log{(1+P\max_i\{h_{Li}^2\})}\right]\\
	&\overset{(b)}{\leq} \frac{k}{2}\log{(1+P\EX\left[\max_i\{h_{Li}^2\}\right])}\\
	&\overset{(c)}{=} \frac{k}{2}\log{\left(1+P\left(2\ln{L}-\ln{\ln{L}}-2\ln{\Gamma\left(\frac{1}{2}\right)}+\frac{\gamma}{2}+o(1)\right)\right)}\\
\end{align*}
where $(a)$ is true since the maximal element in $\b{h}_L$ maximizes the expression and $(b)$ follows from Jensen's inequality. In $(c)$ we used the asymptotic results for the expected value of the maximum value of a $\chi^2$ random vector of dimension $L$ in the limit of large $L$ \cite[Table 3.4.4]{embrechts2013modelling}. 
It can be verified (the computation appears in Appendix \ref{Appendix B}) that the scaling laws indeed behave as $O(\frac{k}{2}\log{\log{L}})$ which completes the proof.
\end{IEEEproof}

\subsection{The value of $k$, completion time and future work}

Up until this point, the number of scheduled users $k$ is assumed as a fixed number. However, it may be optimized and dynamically changed in each transmission in order to provide addition gain to the overall performance of the system. This can be seen in the lower bound given in Theorem \ref{the-Expected sum-rate of scheduling algorithm lower bound} where $k$ constitutes a pre-log factor for the system's sum-rate. We emphasize that one cannot let $k$ be too large (at the order of $L$) and in fact it must satisfy $k<\frac{\delta}{2\sqrt{2\pi}}\sqrt{L}+1$ for the correctness of this bound. Additionally, one should also recall that Theorem \ref{the-Probability for having a unit vector as the maximaizer} implicitly restrict the number of simultaneously transmitting users in order for the CF scheme be applicable. 

Other possible improvement may be realized in the completion time of decoding all messages at the destination. As mentioned earlier, the coefficients vectors form the decoding matrix $\b{A}$ of the linear system of equations to obtain the original $L$ messages. If one let $k$ users to transmit in each transmission phase, he essentially rules the sparseness of this matrix. Note that, although only $k$ users are scheduled for transmission in each phase, the decoding is done simultaneously for all messages so a coefficients vector (at the decoder) in each phase is of dimension $L$ and consist of the coefficients of the $k$ scheduled users and $L-k$ zeroes in the remaining entries. Accordingly, the following question may be asked. How many transmission phases required for complete decoding of all messages as a function of $k$. That is, how many linear combinations the destination must collect until $\b{A}$ has rank $L$ (obviously, $L$ transmission phases must occur). 

One can find resemblance to the known problem of coupons collector, where there are $L$ different coupons which are drawn randomly with replacement. Given this, how many draws are needed on average for the retrieval of all coupons. In our case, each coefficients vector can be considered as a coupon which is innovative or not. Namely, a new vector may increase the rank of the matrix formed by the collected vectors thus far, or it may be linearly dependent. For example, letting $k=1$ means that a single user is scheduled and thus each coefficients vector at the decoder is a unit vector. Since we have $L$ such unit vectors we get exactly the coupons collector problem which needs $O(L\log{L})$ draws on average to obtain all coupons, i.e., $L$ independent coefficients vectors. 

A different variation of the problem described above is considering the case where $k>1$ and in addition, assuming that the coefficients vectors are drawn randomly from the finite field $\mathbb{F}_q^L$ for $q>1$. If $k=L$, i.e., there in no restriction on the vectors, it is not hard to prove that the average number of vectors needed to obtain a matrix $\b{A}$ with rank $L$ is $O(L)$, \cite{eryilmaz2006delay}. Specifically, even if $q=2$ the average number of transmission phases is at most $L+2$ \cite{lucani2009random}.

Considering our scheduling problem, $k<<L$, and thus the received coefficients vectors are restricted to at least $L-k$ zero elements. In addition the elements of the vectors are in $\Z$. We would like to find the value of $k$ for which the average number of transmission phases is $O(L)$. Moreover, we would like to show that if we employ Algorithm \ref{algo-scheduling algorithm}, which ensure high rate for each linear combination by fixing the coefficients vector to be in $\b{a^{\{1\}}}$, this average remains $O(L)$.  We conjecture that in order to fulfil these requirements one needs $k=O(\log{L})$.

\appendices
%%
%% in combination with further \section-commands can be used.
%%%%%%
\section{Proof for the scaling laws of Theorem \ref{the-Expected sum-rate of scheduling algorithm lower bound} }\label{Appendix A}
In order to prove that the scaling laws are $\frac{k}{4}\log{\log{L}}$ we will show that the limit of the division of the lower bound with $\frac{k}{4}\log{\log{L}}$ equals 1 as follows,
\small
\begin{equation}
\begin{aligned}
	&\lim_{L\rightarrow \infty} \frac{\frac{k}{2}  \log^+ \left( k \left(1-\frac{Pku^4}{(u+\delta)^2(1+Pku^2)}\left(1-g(L)\right) \right)\right)^{-1}}{\frac{k}{4}\log{\log{L}}}\\
	&=\lim_{L\rightarrow \infty} \frac{ -2 \log^+ \left( k \left(1-\frac{Pku^4\left(1-g(L)\right)}{(u+\delta)^2(1+Pku^2)} \right)\right)}{\log{\log{L}}}\\
	&=0+\lim_{L\rightarrow \infty} \frac{ -2 \log^+ \left(1-\frac{Pku^4\left(1-g(L)\right)}{(u+\delta)^2(1+Pku^2)}\right)}{\log{\log{L}}},\\
	%&=\lim_{L\rightarrow \infty} \frac{ -2 \log^+ \left(\frac{(u+\delta)^2(1+Pku^2)-Pku^4 \left(1-o(1)\right)}{(u+\delta)^2(1+Pku^2)}\right)}{\log{\log{L}}}\\
	%&=2\lim_{L\rightarrow \infty} \frac{ \log^+ \left( (u+\delta)^2(1+Pku^2) \right) }{\log{\log{L}}} \\
	%&\quad\quad -2\lim_{L\rightarrow \infty}\frac{ \log^+ \left( (u+\delta)^2(1+Pku^2)-Pku^4 \left(1-o(1)\right) \right)}{\log{\log{L}}}\\
	%&\overset{(a)}{=}2\lim_{L\rightarrow \infty} \frac{ \log^+ \left( (\sqrt{2\ln{2\sqrt{L}}})^2(1+Pk(\sqrt{2\ln{2\sqrt{L}}}-\delta)^2) \right) }{\log{\log{L}}} \\
	%&\quad\quad -2\lim_{L\rightarrow \infty}\frac{ \log^+ \left( (\sqrt{2\ln{2\sqrt{L}}})^2(1+Pk(\sqrt{2\ln{2\sqrt{L}}}-\delta)^2)-Pk(\sqrt{2\ln{2\sqrt{L}}}-\delta)^4 \left(1-o(1)\right) \right)}{\log{\log{L}}}
\end{aligned}
\end{equation}
where $g(L)=e^{-\frac{1}{2p(u,\delta)}\frac{(Lp(u,\delta)-(k-1))^2}{L}}$ which we expressed as $o(1)$ in the theorem. We now lower and upper bound this limit to show that both bounds goes to one. Let us start with the upper bound.
\begin{align*}
	&=\lim_{L\rightarrow \infty} \frac{ -2 \log^+ \left(1-\frac{Pku^4\left(1-g(L)\right)}{(u+\delta)^2(1+Pku^2)}\right)}{\log{\log{L}}}\\
	&\leq \lim_{L\rightarrow \infty} \frac{ -2 \log^+ \left(1-\frac{Pku^4}{(u+\delta)^2(1+Pku^2)}\right)}{\log{\log{L}}}\\
	&\leq \lim_{L\rightarrow \infty} \frac{ -2 \log^+ \left(1-\frac{u^2}{(u+\delta)^2}\right)}{\log{\log{L}}}\\
	&\leq \lim_{L\rightarrow \infty} \frac{ -2 \log^+ \left(\frac{2\delta u+\delta^2}{(u+\delta)^2}\right)}{\log{\log{L}}}\\
	&\leq \lim_{L\rightarrow \infty} \frac{ -2 \log^+ \left(\frac{\delta u+\delta^2}{(u+\delta)^2}\right)}{\log{\log{L}}}\\
	&\overset{(a)}{=} \lim_{L\rightarrow \infty} \frac{ -2 \log^+ \left(\frac{\delta\sqrt{2\log{2\sqrt{L}}} }{2\log{2\sqrt{L}}}\right)}{\log{\log{L}}}\\
	&= \lim_{L\rightarrow \infty} \frac{ \log^+ \left(2\log{2\sqrt{L}} \right)}{\log{\log{L}}}\\
	&= \lim_{L\rightarrow \infty} \frac{ \log\log{4L}}{\log{\log{L}}}=1.
\end{align*}

In $(a)$ we set $u=\sqrt{2\ln{2\sqrt{L}}}-\delta$. The lower bound is as follows,
\begin{align*}
	&=\lim_{L\rightarrow \infty} \frac{ -2 \log^+ \left(1-\frac{Pku^4\left(1-g(L)\right)}{(u+\delta)^2(1+Pku^2)}\right)}{\log{\log{L}}}\\
	&\geq \lim_{L\rightarrow \infty} \frac{ -2 \log^+ \left(1-\frac{u^4\left(1-g(L)\right)}{(u+\delta)^2(1+u^2)}\right)}{\log{\log{L}}}\\
	&= \lim_{L\rightarrow \infty} \frac{ 2 \log^+ \left(\frac{(u+\delta)^2(1+u^2)}{(u+\delta)^2(1+u^2)-u^4\left(1-g(L)\right)}\right)}{\log{\log{L}}}\\
	&\geq \lim_{L\rightarrow \infty} \frac{ 2 \log^+ \left(\frac{(u+\delta)^2u^2}{(u+\delta)^2+2u^3\delta+u^2\delta^2+u^4g(L)}\right)}{\log{\log{L}}}\\
	&\geq \lim_{L\rightarrow \infty} \frac{ 2 \log^+ \left(\frac{(u+\delta)^2u^2}{(u+\delta)^2+2(u+\delta)^3\delta+(u+\delta)^2\delta^2+(u+\delta)^4g(L)}\right)}{\log{\log{L}}}\\
	&= \lim_{L\rightarrow \infty} \frac{ 2 \log^+ \left(\frac{u^2}{1+2(u+\delta)\delta+\delta^2+(u+\delta)^2g(L)}\right)}{\log{\log{L}}}\\
	&\overset{(a)}{=} \lim_{L\rightarrow \infty} \frac{ 2 \log^+ \left(\frac{\left(\sqrt{2\ln{2\sqrt{L}}}-\delta\right)^2}{1+2\delta\sqrt{2\ln{2\sqrt{L}}}+\delta^2+g(L)2\ln{2\sqrt{L}}}\right)}{\log{\log{L}}}\\
	&\geq \lim_{L\rightarrow \infty} \frac{ 2 \log^+ \left(\frac{2\ln{2\sqrt{L}}-2\delta\sqrt{2\ln{2\sqrt{L}}}}{1+2\delta\sqrt{2\ln{2\sqrt{L}}}+\delta^2+g(L)2\ln{2\sqrt{L}}}\right)}{\log{\log{L}}}\\
	&= \lim_{L\rightarrow \infty} \frac{ 2 \log^+ \left(\sqrt{2\ln{2\sqrt{L}}}-2\delta\right)}{\log{\log{L}}}\\
	&\quad\quad  - \lim_{L\rightarrow \infty} \frac{ 2 \log^+ \left(\frac{1+\delta^2}{\sqrt{2\ln{2\sqrt{L}}}}+2\delta+g(L)\sqrt{2\ln{2\sqrt{L}}}\right)}{\log{\log{L}}}\\
	&\geq \lim_{L\rightarrow \infty} \frac{  \log^+ \left(2\ln{2\sqrt{L}}-2\delta\sqrt{2\ln{2\sqrt{L}}}\right)}{\log{\log{L}}}\\
	&\quad\quad  - \lim_{L\rightarrow \infty} \frac{ 2 \log^+ \left(\frac{2\delta^2}{\sqrt{2\ln{2\sqrt{L}}}}+2\delta+g(L)2\ln{2\sqrt{L}}\right)}{\log{\log{L}}}\\
	&\geq \lim_{L\rightarrow \infty} \frac{  \log^+ \left(2\ln{2\sqrt{L}}-2\delta2\ln{2\sqrt{L}}\right)}{\log{\log{L}}}\\
	&\quad\quad  - \lim_{L\rightarrow \infty} \frac{ 2 \log^+ \left(\delta^2+\delta+g(L)\ln{2\sqrt{L}}\right)}{\log{\log{L}}}\\
	&\geq \lim_{L\rightarrow \infty} \frac{  \log^+ \left(\ln{2\sqrt{L}}\right)}{\log{\log{L}}}\\
	&\quad\quad  - \lim_{L\rightarrow \infty} \frac{ 2 \log^+ \left(\left(\delta^2+\delta+1\right)\left(g(L) \ln{2\sqrt{L}}+1\right)\right)}{\log{\log{L}}}\\
	&= 1 - \lim_{L\rightarrow \infty} \frac{ 2 \log^+ \left(1+g(L) \ln{2\sqrt{L}}\right)}{\log{\log{L}}}\\
	&\geq 1 - \lim_{L\rightarrow \infty} \frac{ 2 \log^+ \left(1+g(L)L\right)}{\log{\log{L}}}\\
	&\overset{(b)}{=} 1 - \lim_{L\rightarrow \infty} \frac{ 2 \log^+ \left(1+e^{-\frac{\sqrt{2\pi}\sqrt{L}}{\delta}{\left(  \frac{\delta}{2\sqrt{2\pi}}-\frac{(k-1)}{\sqrt{L}}\right)^2}}L\right)}{\log{\log{L}}}\\
	&= 1 - \lim_{L\rightarrow \infty} \frac{ 2 \log^+ \left(1+e^{-\frac{\sqrt{L}\delta}{4\sqrt{2\pi}}}e^{(k-1)}e^{-\frac{\sqrt{2\pi}(k-1)^2}{\delta\sqrt{L}}}L\right)}{\log{\log{L}}}\\
	&\geq 1 - \lim_{L\rightarrow \infty} \frac{ 2 \log^+ \left(1+e^{-\frac{\sqrt{L}\delta}{4\sqrt{2\pi}}}e^{(k-1)}L\right)}{\log{\log{L}}}\\
	&= 1 - \lim_{L\rightarrow \infty} \frac{ 2 \log^+ \left(1+e^{-\frac{\sqrt{L}\delta}{4\sqrt{2\pi}}}L\right)}{\log{\log{L}}}\\
	&\overset{(c)}{=} 1 - 2\lim_{L\rightarrow \infty} \frac{ \left(2 - \frac{\delta}{4\sqrt{2\pi}}\sqrt{L}\right) L \log L}{2\left(e^{\frac{\sqrt{L}\delta}{4\sqrt{2\pi}}}+L\right)}\\
	&=1
	\end{align*}
\normalsize
In $(a)$ we set $u=\sqrt{2\ln{2\sqrt{L}}}-\delta$, in $(b)$ we set $g(L)$ with its expression with $p(u,\delta)$ as in \eqref{equ-probability p that will go to one} and in $(c)$ we used L'Hospital's rule which completes the proof.

\section{Proof for the scaling laws of Theorem \ref{the-Expected sum-rate of scheduling algorithm upper bound} }\label{Appendix B}
In order to prove that the scaling laws are $\frac{k}{2}\log{\log{L}}$ we will show that the limit of the division of the lower bound with $\frac{k}{2}\log{\log{L}}$ equals 1 as follows,
\small
\begin{equation}
\begin{aligned}
	&\lim_{L\rightarrow \infty} \frac{\frac{k}{2}\log{\left(1+P\left(2\ln{L}-\ln{\ln{L}}-2\ln{\Gamma\left(\frac{1}{2}\right)}+\frac{\gamma}{2}\right)\right)}}{\frac{k}{2}\log{\log{L}}}\\
	&\lim_{L\rightarrow \infty} \frac{\log{\left(1+P\left(2\ln{L}-\ln{\ln{L}}-2\ln{\Gamma\left(\frac{1}{2}\right)}+\frac{\gamma}{2}\right)\right)}}{\log{\log{L}}}\\
\end{aligned}
\end{equation}
We now lower and upper bound this limit to show that both bounds goes to one. Let us start with the upper bound.
\begin{equation}
\begin{aligned}
	&\lim_{L\rightarrow \infty} \frac{\log{\left(1+P\left(2\ln{L}-\ln{\ln{L}}-2\ln{\Gamma\left(\frac{1}{2}\right)}+\frac{\gamma}{2}\right)\right)}}{\log{\log{L}}}\\
	&\leq \lim_{L\rightarrow \infty} \frac{\log{\left(P+P\left(2\ln{L}-\ln{\ln{L}}-2\ln{\Gamma\left(\frac{1}{2}\right)}+\frac{\gamma}{2}\right)\right)}}{\log{\log{L}}}\\
	&\leq \lim_{L\rightarrow \infty} \frac{\log{\left(P\left(1+2\ln{L}-\ln{\ln{L}}-2\ln{\Gamma\left(\frac{1}{2}\right)}+\frac{\gamma}{2}\right)\right)}}{\log{\log{L}}}\\
	&\leq \lim_{L\rightarrow \infty} \frac{\log{\left(1+2\ln{L}-\ln{\ln{L}}-2\ln{\Gamma\left(\frac{1}{2}\right)}+\frac{\gamma}{2}\right)}}{\log{\log{L}}}\\
	&\leq \lim_{L\rightarrow \infty} \frac{\log{\left(1+2\ln{L}+\frac{\gamma}{2}\right)}}{\log{\log{L}}}=1.
\end{aligned}
\end{equation}
The lower bound is as follows,
\begin{equation}
\begin{aligned}
	&\lim_{L\rightarrow \infty} \frac{\log{\left(1+P\left(2\ln{L}-\ln{\ln{L}}-2\ln{\Gamma\left(\frac{1}{2}\right)}+\frac{\gamma}{2}\right)\right)}}{\log{\log{L}}}\\
	&\geq \lim_{L\rightarrow \infty} \frac{\log{\left(2\ln{L}-\ln{\ln{L}}-2\ln{\Gamma\left(\frac{1}{2}\right)}+\frac{\gamma}{2}\right)}}{\log{\log{L}}}\\
	&\geq \lim_{L\rightarrow \infty} \frac{\log{\left(2\ln{\ln{L}}-\ln{\ln{L}}-2\ln{\Gamma\left(\frac{1}{2}\right)}+\frac{\gamma}{2}\right)}}{\log{\log{L}}}\\
	&\geq \lim_{L\rightarrow \infty} \frac{\log{\left(\ln{\ln{L}}-2\ln{\Gamma\left(\frac{1}{2}\right)}\right)}}{\log{\log{L}}}=1.
\end{aligned}
\end{equation}

%%%%%%
%% To balance the columns at the last page of the paper use this
%% command:
%%
%\enlargethispage{-1.2cm} 
%%
%% If the balancing should occur in the middle of the references, use
%% the following trigger:
%%
%%\IEEEtriggeratref{3}
%%
%% which triggers a \newpage (i.e., new column) just before the given
%% reference number. Note that you need to adapt this if you modify
%% the paper.  The "triggered" command can be changed if desired:
%%
%\IEEEtriggercmd{\enlargethispage{-20cm}}
%%
%%%%%%

%%%%%%
%% References:
%% We recommend the usage of BibTeX:
%%
%\bibliographystyle{IEEEtran}
%\bibliography{definitions,bibliofile}
%%
%% where we here have assume the existence of the files
%% definitions.bib and bibliofile.bib.
%% BibTeX documentation can be obtained at:
%% http://www.ctan.org/tex-archive/biblio/bibtex/contrib/doc/
%%%%%%

\bibliographystyle{IEEEtran}
\bibliography{../main_document/Bibliography}

\end{document}